%% file: main.tex
\documentclass[sigconf,preprint]{acmart}
\setcopyright{none}
\acmConference[]{}{}{}
\renewcommand\footnotetextcopyrightpermission[1]{}

\usepackage{listings}
\input{lstlistings_python}

\newcommand{\code}[1]{\texttt{#1}}

\definecolor{changedcolor}{rgb}{0,0,.5}

\begin{document}

\author{Michael Pradel}
\email{michael@binaervarianz.de}
\affiliation{
\institution{University of Stuttgart}
\country{Germany}
}

\author{Satish Chandra}
\email{schandra@acm.org}
\affiliation{
\institution{Facebook}
\country{USA}
}

\title{Neural Software Analysis}

\begin{abstract}
Many software development problems can be addressed by program analysis tools, which traditionally are based on precise, logical reasoning and heuristics to ensure that the tools are practical.
Recent work has shown tremendous success through an alternative way of creating developer tools, which we call \emph{neural software analysis}.
The key idea is to train a neural machine learning model on numerous code examples, which, once trained, makes predictions about previously unseen code.
In contrast to traditional program analysis, neural software analysis naturally handles fuzzy information, such as coding conventions and natural language embedded in code, without relying on manually encoded heuristics.
This article gives an overview of neural software analysis, discusses when to (not) use it, and presents three example analyses.
The analyses address challenging software development problems: bug detection, type prediction, and code completion.
The resulting tools complement and outperform traditional program analyses, and are used in industrial practice.
\end{abstract}

\maketitle
\pagestyle{plain}

\section{Introduction}

Software is increasingly dominating the world.
The huge demand for more and better software is turning tools and techniques for software developers  into an important factor toward a productive economy and strong society.
Such tools aim at making developers more productive by supporting them through (partial) automation in various development tasks.
For example, developer tools complete partially written code, warn about potential bugs and vulnerabilities, find code clones, or help developers search through huge code bases.

The conventional way of building developer tools is program analysis based on precise, logical reasoning.
Such traditional program analysis is deployed in compilers and many other widely used tools.
Despite its success, there are many problems that traditional program analysis can only partially address.
The reason is that practically all interesting program analysis problems are undecidable, i.e.,
giving answers guaranteed to be precise and correct is impossible for non-trivial programs.
Instead, program analysis must approximate the behavior of the analyzed software, often with the help of carefully crafted heuristics.

Crafting effective heuristics is difficult, especially because the correct analysis result often depends on uncertain information, e.g., natural language information or common coding conventions, that is not amenable to precise, logic-based reasoning.
Fortunately, software is written by humans and hence follows regular patterns and coding idioms, similar to natural language~\cite{Hindle2012}.
For example, developers commonly call a loop variable \code{i} or \code{j}, and most developers 
prefer a \code{for}-loop over a \code{while}-loop when iterating through a sequential data structure.
This ``naturalness'' of software has motivated research on machine learning-based software analysis that exploits the regularities and conventions of code~\cite{Raychev2015,Allamanis2018}.

Over the past years, deep neural networks have emerged as a powerful technique to reason about uncertain data and to make probabilistic predictions.
Can software be considered ``data'' for neural networks?
This article answers the question with a confident ``yes''.
We present a recent stream of research on what we call \emph{neural software analysis} -- an alternative take at program analysis based on neural machine learning models that reason about software.

The remainder of this article starts by defining criteria for when to use neural software analysis based on when it is likely to complement or even outperform traditional program analysis.
We then present a conceptual framework that shows how neural software analyses are typically built, and illustrate it with a series of examples, three of which we describe in more detail.
The example analyses address common development problems, such as bug detection or code completion, and are already used by practitioners, despite the young age of the field.
Finally, we discuss open challenges and give an outlook into promising directions for future work on neural software analysis.

\section{When to Use Neural Software Analysis}
\label{sec:dimensions}

\begin{figure}
    \centering
    \includegraphics[width=.9\linewidth]{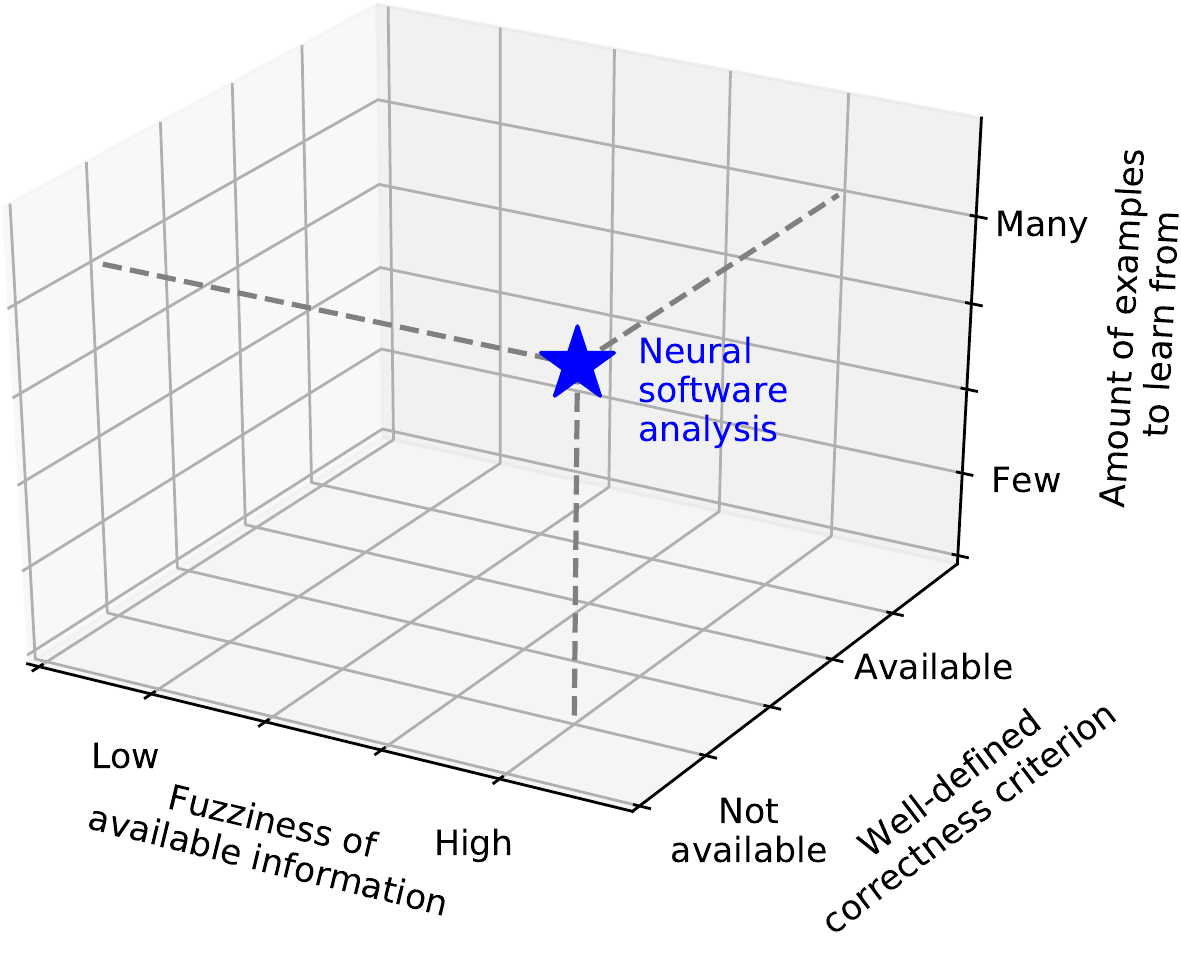}
    \caption{Three dimensions to determine whether to use neural software analysis.}
    \label{fig:dimensions}
\end{figure}

In principle, practically all program analysis problems can be formulated in a traditional, logic reasoning-based way, as well as a data-driven, learning-based way, such as neural software analysis.
The following describes conditions where neural software analysis is most suitable, and likely to outperform traditional, logic-based program analysis.
See Figure~\ref{fig:dimensions} for a visual illustration.

\paragraph{Dimension 1: Fuzziness of the available information}
Traditional program analysis is based on drawing definite conclusions from exact input information.
However, such precise, logical reasoning often fails to represent uncertainties.
Neural software analysis instead is able to handle fuzzy inputs given to the analysis, e.g., natural language embedded in code.
The more fuzzy the available information is, the more likely it is that neural software analysis is suitable.
The reason is that neural models identify patterns while allowing for an imprecise description of these patterns.
For example, instead of relying on a strict rule of the form ``If the code has property A, then B holds'', as traditional program analysis would use, neural software analysis learns fuzzy rules of the form ``If the code is similar to pattern A, then B is likely to hold''.

\paragraph{Dimension 2: Well-defined correctness criterion}
Some program analysis problems have a well-defined correctness criterion, or specification, which precisely describes when an answer offered by an analysis is what the human wants, without checking with the human.
For example, this is the case for test-guided program synthesis,
where the provided test cases specify when an answer is correct, or for an analysis that type checks a program, where the rules of a type system define when a program is guaranteed to be type-safe.
In contrast, many other analysis problems do not offer the luxury of a well-defined correctness criterion.
For these problems, a human developer ultimately decides whether the answer by the analysis fits the developer's needs, typically based on whether the analysis successfully imitates what a developer would do.
For example, such problems include code search based on a natural language query, code completion based on partially written code, or predicting whether a code change risks causing bugs.
Neural software analysis often outperforms traditional analysis for problems that lack a well-defined correctness criterion.
The reason is that addressing such ``specification-free'' problems is ultimately a matter of finding suitable heuristics, a task at which learned models are very effective.

\paragraph{Dimension 3: Amount of examples to learn from}
Neural models are data-hungry, and hence, neural software analysis works best if there are plenty of examples to learn from.
Typically, training an effective neural model requires at least several thousands of examples.
These examples can come in various forms, e.g., code snippets extracted from a large code corpus.
Some neural software analyses do not only extract examples from the code as-is, but also modify the code to create examples of an otherwise underrepresented class, e.g., buggy code.

\section{A Conceptual Framework for Neural Software Analysis}

\begin{figure}
    \centering
    \includegraphics[width=\linewidth]{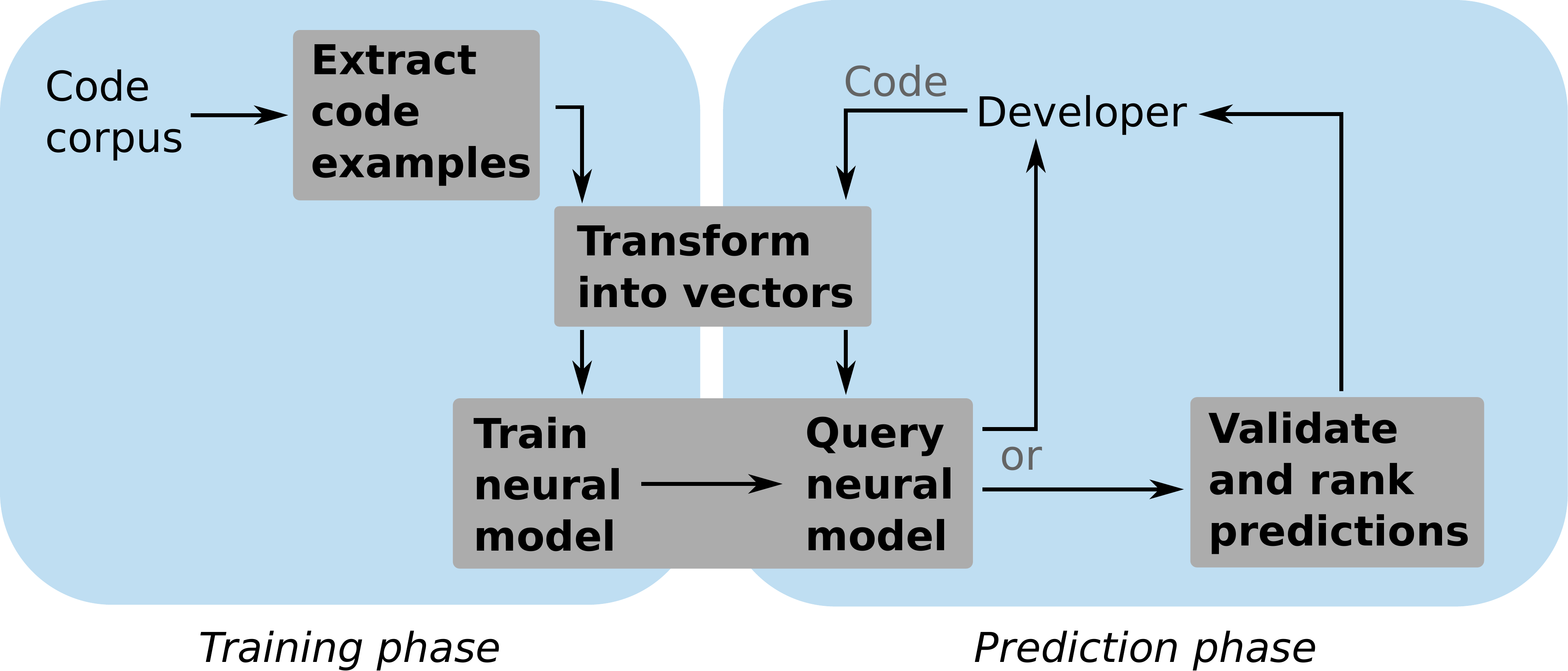}
    \caption{Typical components of a neural software analysis.}
    \label{fig:framework}
\end{figure}

Many neural software analyses have an architecture that consists of five components (Figure~\ref{fig:framework}).
Given a code corpus to learn from, the first component \emph{extracts code examples} suitable for the problem the analysis tries to address.
These code examples are \emph{transformed into vectors} -- either based on an intermediate representation known from compilers, such as token sequences or abstract syntax trees, or a novel code representation developed specifically for learning-based analysis.
Next, the examples serve as training data to \emph{train a neural model}.
The first three steps all happen during a training phase that is performed only once, before the analysis is deployed to developers.
After training, the analysis enters the prediction phase, where a developer \emph{queries the neural model} with previously unseen code examples.
The model yields predictions for the given query, which are either given directly to the developer or go through an optional \emph{validation and ranking} component.
Validation and ranking sometimes rely on a traditional program analysis, combining the strengths of both neural and logic-based reasoning.
The remainder of this section discusses these components in more detail.

\subsection{Extracting Code Examples}

\paragraph{Lightweight static analysis}
To extract code examples to learn from, most neural software analyses build on a lightweight static analysis.
Such a static analysis reuses standard tools and libraries available for practically any programming language, e.g., a tokenizer, which splits the program code into tokens, or a parser, which transforms the program code into an abstract syntax tree (AST).
These tools are readily available as part of an IDE or a compiler.
Building on such a lightweight, standard static analysis, instead of resorting to more sophisticated static analyses, is beneficial in two ways.
First, it ensures that the neural software analysis scales well to large code corpora, which are typically used for effective training.
Second, it makes it easy to port a neural software analysis developed for one language to another language.

\paragraph{Obtaining labeled examples}
Practically all existing neural software analyses use some form of supervised learning.
They hence require labeled code examples, i.e., code examples that come with the desired prediction result, so that the neural model can learn from it.
How to obtain such labeled examples depends on the specific task an analysis is addressing.
For example, an analysis that predicts types can learn from existing type annotations~\cite{Hellendoorn2018,icse2019,fse2020}, and an analysis that predicts code edits can learn from edit histories, e.g., documented in a version control system~\cite{Tufano2019,Dinella2020}.
Because large amounts of labeled examples are a prerequisite for effective supervised learning, what development tasks receive most attention by the neural software analysis community is partially driven by the availability of sufficiently large annotated datasets.

\subsection{Representing Software as Vectors}
\label{sec:vectors}

Since neural models reason about vectors of numbers, the perhaps most important design decision of a neural software analysis is how to turn the code examples extracted in the previous step into vectors.
We discuss two aspects of this step:
(1) How to represent the basic building blocks of code, i.e., individual code tokens.
(2) How to compose representations of the basic building blocks into representations of larger snippets of code, e.g., statements or functions.

\paragraph{Representing Code Tokens}
Any technique for representing code as vectors faces the question of how to map the basic building blocks of the programming language into vectors.
Most neural software analyses address the token representation challenge in one of two ways.
One approach is to abstract away all non-standard tokens in the code, e.g., by abstracting variable names into \code{var1}, \code{var2}, etc.~\cite{Gupta2017,Tufano2019}.
While this approach effectively reduces the vocabulary size, it also discards potentially useful information.
The other approach maps each token into an embedding vector of a fixed size.
The goal here is to represent semantically similar tokens, e.g., the two identifiers \code{len} and \code{size}, with similar vectors~\cite{icse2021}.
To obtain such an embedding, some analyses train a token embedding before training the neural model that addresses the main task, and then map each token to a vector using the pre-trained embedding~\cite{oopsla2018-DeepBugs,Kanade2020}.
Alternatively, some analyses learn an embedding function jointly with the overall neural model, essentially making the task of handling the many different identifiers part of the overall optimization task that the machine learning model addresses~\cite{Allamanis2018b}.

A key challenge is the fact that the vocabulary of identifiers that developers can freely choose, e.g., variable and function names, grows in an apparently linear fashion when new projects are added to a code corpus~\cite{Karampatsis2020a}.
The reason is that developers come up with new terminology and conventions for different applications and application domains, leading to multiple millions of different identifiers in a corpus of only a few thousand projects.
The most simple approach to handle the vocabulary problem is to fix the vocabulary to the, say, 10,000 most common tokens, while representing all other, out-of-vocabulary tokens with as a special ``unknown'' vector.
More sophisticated techniques split tokens into subwords, e.g., \code{writeFile} into ``write'' and ``file'', represent each subword individually, and then compose subword vectors into the representation of a full token.
To split tokens into subwords, neural software analyses can rely on conventions~\cite{Allamanis2016} or compression algorithms that compute a fixed-size set of those subwords that occur most frequently~\cite{Karampatsis2020a}.

\paragraph{Representing Snippets of Code}
How to turn snippets of source code, e.g., the code in a statement or function, into a vector representation has received lots of attention by researchers and practitioners recently.
The many proposed techniques can be roughly summarized into two groups.
Both of them rely on some way of mapping the most elementary building blocks of code into vectors, as described above.
On the one hand, there are techniques that turn a code snippet into one or more sequences of vectors.
The most simple, yet quite popular and effective, technique~\cite{Gupta2017,Hellendoorn2018,Tufano2019} starts from the sequence of code tokens and maps each token into a vector.
E.g., a code snippet \code{x=true;} would be mapped into a sequence of four vectors that represent ``x'', ``='', ``true'', and ``;'', respectively.
Instead of viewing code as a flat sequence of tokens, other techniques leverage the fact that code, in contrast to, e.g., natural language, has a well-defined and unambiguous structure~\cite{Mou2016,Alon2019}.
Such techniques typically start from the AST of a code snippet and extract one or more paths through the tree, mapping each node in a path to a vector.
E.g., the popular code2vec technique~\cite{Alon2019} extracts many such AST paths, each connecting two leaves in the tree.

On the other hand, several techniques represent code snippets as graphs of vectors~\cite{Allamanis2018b,Dinella2020,Wei2020}.
These graphs are typically based on ASTs, possibly augmented with additional edges that represent data flow, control flow, and other relations between code elements that can be computed by a traditional static analysis.
Given such a graph, these techniques map each node into a vector, yielding a graph of vectors.
The main advantage of graph-based vector representations of code is that they provide the rich structural and semantic information available for code to the neural model.
On the downside, rich graph representations of code are less portable across programming languages, and graph-based neural models tend to be computationally more expensive than sequence-based models.

\subsection{Neural Models of Software}

\begin{figure*}
    \centering
    \includegraphics[width=.85\linewidth]{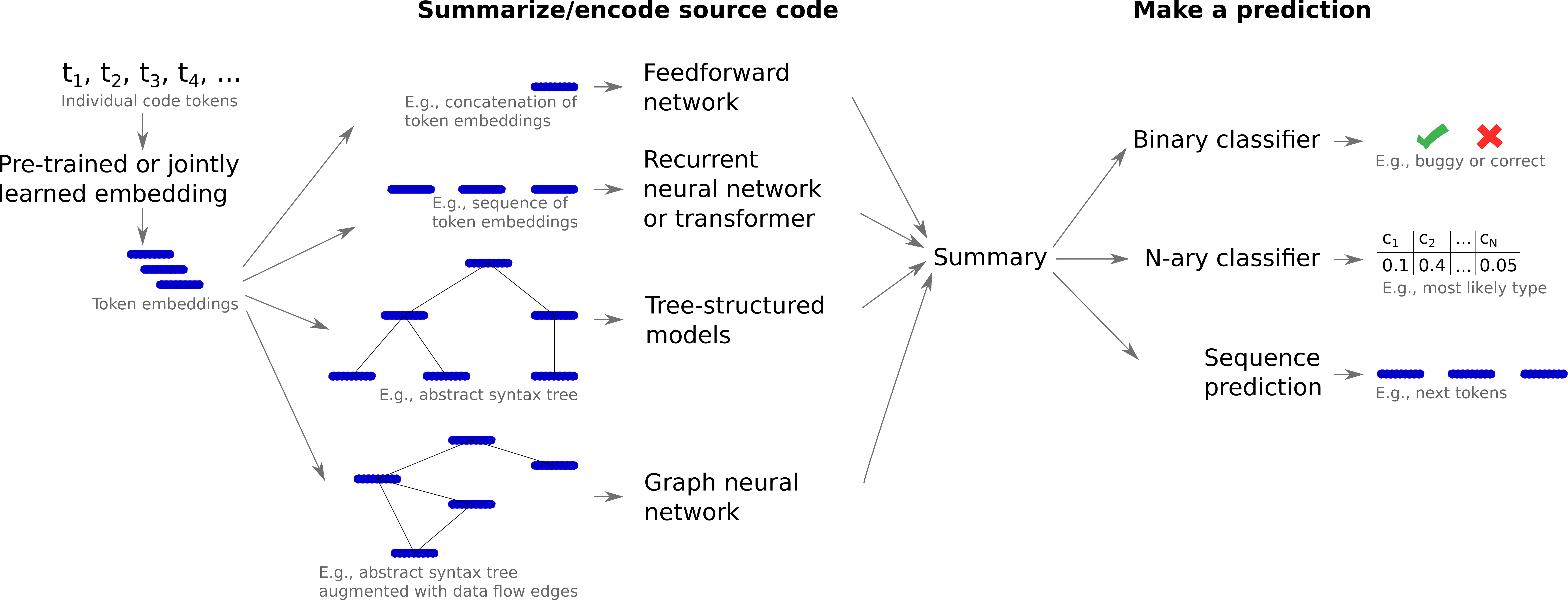}
    \caption{Neural components popular for analyzing software. In principle, any of the summarization components on the left can be freely combined with any of the prediction components on the right.}
    \label{fig:neural_models}
\end{figure*}

Once the source code is represented as vectors, the next step is to feed these vectors into a machine learning model.
Neural models of software typically consist of two parts.
One part summarizes (or encodes) the given source code into a more compact representation, e.g., a single vector.
Given this summary, the other part then makes a prediction about the code.
Figure~\ref{fig:neural_models} shows popular neural components used in these two parts, which we discuss in more detail in the following.
Each neural component essentially corresponds to a learned function that maps some input to some output.

\paragraph{Summarizing the code source}
The simplest way of summarizing code is to concatenate the vectors that describe it and map the concatenation into a shorter vector using a feedforward network.
For code that is represented as a sequence of vectors, e.g., each representing a token, recurrent neural networks or transformers~\cite{Vaswani2017} can summarize them. 
While the former traverses the sequence once, the latter iteratively pays attention to specific elements of the sequence, where the decision what to pay attention to is learned.
Another common neural component are tree-structured models that summarize a tree representation of code, e.g., the AST of a code snippet~\cite{Mou2016,Alon2019}.
Finally, the perhaps most sophisticated of today's techniques are graph neural networks (GNNs)~\cite{Allamanis2018b}, which operate on a graph representation of the code.
Given a graph of initial vector representations of each node, a GNN repeatedly updates nodes based on the current representations of the neighboring nodes, effectively propagating information across code elements.

\paragraph{Making a prediction}
Neural models of code are almost exclusively classification models, which is motivated by the fact that most information associated with programs is discrete.
One common prediction component is a binary classifier, e.g., to predict whether a piece of code is correct or buggy~\cite{oopsla2018-DeepBugs,Li2018a}.
In a similar vein, an $N$-ary classifier predicts which class(es) out of a fixed set of $N$ classes the code belongs to, e.g., predicting the type of a variable out of a set of common types~\cite{Hellendoorn2018,icse2019,fse2020}.
Such classifiers output a probability distribution over the available classes, which the model obtains by passing the unscaled outputs (\emph{logits}) of earlier layers of the network through the softmax function.
Beyond flat classification, some neural code analyses predict sequences of vectors, e.g., a sequence of code tokens~\cite{kim-icse-2021}, how to edit a given piece of code~\cite{Tufano2019}, or a natural language description of the code~\cite{Allamanis2016}.
A common neural model for such tasks is an encoder-decoder model, which combines a sequence encoder with a decoder that predicts a sequence.
In these models, the decoder's output at each step is a probability distribution, e.g., across possible source-level tokens, similar to a flat classifier.

\paragraph{Training and querying}
Training and querying neural models of software works just like in any other domain:
The parameters, called weights and biases, of the various functions that control a model's behavior are optimized to fit examples of inputs and expected outputs.
For example, for an $N$-ary classification model, training will compare the predicted probability distribution to the desired distribution via a loss function, e.g., cross entropy, and then try to minimize the loss through stochastic gradient descent.
Once successfully trained, the model makes predictions about previously unseen software by generalizing from the examples seen during training.  



\subsection{Validation and Ranking}
\label{sec:validation ranking}

The final component of some neural software analyses validates and ranks the predictions made by the neural model.
Validation is optional and can come in various, task-specific forms.
For example, the predictions of a model that suggests new code can be filtered for syntactic correctness using the grammar of the programming language~\cite{Li2020a}.
As another example, the predictions of a model that suggests types can be validated using a type checker, to ensure that only type-correct suggestions are provided to the user~\cite{fse2020}.

For ranking, one option is to use the numeric vectors predicted by the model to identify the predictions the model is most confident about.
In a classification model that uses the softmax function, many approaches interpret the predicted likelihood of classes as a probability distribution, and then rank the classes by their predicted probability~\cite{Hellendoorn2018,fse2020}.
In an encoder-decoder model, beam search is commonly used to obtain the $k$ most likely predicted outputs by keeping the $k$ overall most likely predictions while decoding a complex output, such as a sequence of code tokens~\cite{Tufano2019}.


\section{Examples of Neural Software Analysis Tools}

\begin{table}[]
    \centering
    \caption{Examples of neural software analyses.}
    \label{tab:analyses}
    \begin{tabular}{@{}p{11em}p{15em}@{}}
        \toprule
        Analysis problem & Neural software analyses \\
        \midrule
        Bug detection
        & DeepBugs~\cite{oopsla2018-DeepBugs} \\
        & VarMisuse using GGNN~\cite{Allamanis2018b} \\
        & VulDeePecker~\cite{Li2018a} \\
        & GREAT~\cite{Hellendoorn2020} \\
        
        Program repair
        & DeepFix~\cite{Gupta2017} \\
        & SequenceR~\cite{Chen2019} \\
        & Hoppity~\cite{Dinella2020} \\
        & Graph2Diff~\cite{DBLP:conf/icse/TarlowMRCMSA20} \\
        
        Code captioning
        & Code summarization~\cite{Allamanis2016}\\
        & Code2Seq~\cite{Alon2019a}\\
        
        Type prediction
        & DeepTyper~\cite{Hellendoorn2018}\\
        & NL2Type~\cite{icse2019} \\
        & LambdaNet~\cite{Wei2020} \\
        & TypeWriter~\cite{fse2020} \\
        
        Code synthesis
        & RobustFill~\cite{RobustFill} \\
        & Autopandas~\cite{AutoPandas} \\
    
        Code completion
        & \citet{li2018code-rnn-attn}\\
        & IntelliCompose~\cite{IntelliCompose}\\
        & \citet{Karampatsis2020a} \\
        & TravTrans~\cite{kim-icse-2021}\\
        
        Clone detection
        & \citet{White2016} \\
    
        Reverse engineering
        & DIRE~\cite{Lacomis2019} \\
        & \citet{David2020} \\
        
        Code search
        & Neural code search~\cite{Sachdev2018} \\
        & Deep code search~\cite{Gu2018} \\
        
        Code-comment matching & \citet{DBLP:conf/aaai/PanthaplackelGM20} \\
    
        \bottomrule
    \end{tabular}
\end{table}

Given the conceptual framework for neural software analysis, we now discuss concrete example analyses.
As a broader overview, Table~\ref{tab:analyses} shows selected analysis problems that are suitable for neural software analysis, along with some representative tools addressing these problems.
\citet{Allamanis2018} provide an extensive survey of many more analyses.
The tools address a diverse set of problems, ranging from classification tasks, such as bug detection or type prediction, over generation tasks, such as code completion and code captioning, to retrieval tasks, such as code search.

\begin{figure}
    \centering
    \begin{lstlisting}
class Board:
  /*#\HL#*/ TypeWriter infers the function signature /*#\HLoff#*/
  /*#\HL#*/ (Board, int, int, str) -> Bool /*#\HLoff#*/
  def mark_point(self, x, y, player_name):
    """
      Marks the given point on the board
      as chosen by the given player.
      Returns whether the move gives the player
      three marked fields in a row.
    """
    self.field[x][y] = player_name
    has_three_in_a_row = False
    ... # compute whether the player has won
    return has_three_in_a_row /*#\label{line:return}#*/
  
  def show_winner(self, player_name):
    ...
    
while not game_done: /*#\label{line:loop_start}#*/
  active_player = ...
  x = ...
  y = ...
  /*#\HL#*/ DeepBugs warns about a bug here: /*#\HLoff#*/
  has_won = board.mark_point(y, x, active_player) /*#\label{line:bug}#*/
  if has_won:
    # notify player
    game_done = True
    board.???/*#\label{line:completion}#*/ /*#\HL#*/ Neural model suggests completions here /*#\HLoff#*/ /*#\label{line:loop_end}#*/
    \end{lstlisting}
    \caption{Python implementation of a tic-tac-toe game.}
    \label{fig:example}
\end{figure}

The remainder of this section discusses three concrete examples of neural software analysis in some more detail.
To illustrate the example analyses, we use a small Python code example (Figure~\ref{fig:example}).
The example is part of an implementation of a tic-tac-toe game and centers around a class \code{Board} that represents the 3x3 grid the game is played on.
The main loop of the game (lines~\ref{line:loop_start} to~\ref{line:loop_end}) implements the turns that players take.
In each turn, a player marks a point on the grid, until one of the players marks three points in a row or until the grid is completely filled.

\begin{table*}[]
    \centering
    \caption{Three neural software analyses and how they map onto the conceptual framework in Figure~\ref{fig:framework}.}
    \label{tab:detailed examples}
    \begin{tabular}{@{}p{8em}|p{15em}p{15em}p{15em}@{}}
        \toprule
          & \multicolumn{3}{c}{Neural software analysis} \\
        \cmidrule{2-4}
        Component of conceptual framework & Bug detection (DeepBugs) & Type prediction (TypeWriter) & Code completion\\
        \midrule
        Code corpus
        & JavaScript (68M lines of open-source code)
        & Python (2.7M lines of open-source code and a larger commercial corpus)
        & Python (16M lines of open-source code) 
        \\
        Extraction of code examples
        & Code snippets as-is and with artificially introduced bugs
        & Functions with their parameter and return types
        & Code token sequences, offset by one for next token prediction
        \\
        Transformation into vectors
        & Concatenation of token embeddings and context information
        & Token embeddings for code, word embeddings for comments
        & End-to-end learned token embeddings for code
        \\
        Neural model &
        Simple feedforward model
        & Hierarchical model built from several recurrent neural networks
        & Encoder (bi-directional LSTM) and decoder (LSTM)
        \\
        Validation and ranking
        & Rank warnings by predicted probability that code is buggy
        & Search and validate correct types with type checker
        & Rank output tokens by probability that it is the next token
        \\
        \bottomrule
    \end{tabular}
\end{table*}

We use this example to illustrate three neural analyses summarized in Table~\ref{tab:detailed examples}.
The analyses are useful for finding a bug in the example (Section~\ref{sec:DeepBugs}), predicting the type of a function (Section~\ref{sec:TypeWriter}), and completing the example's code (Section~\ref{sec:code completion}), respectively.

\subsection{Learning to Find Bugs}
\label{sec:DeepBugs}

DeepBugs~\cite{oopsla2018-DeepBugs} is a neural software analysis that tackles a continuously important problem in software development -- the problem of finding bugs.
While there is a tremendous amount of work on traditional, logic-based analyses to find bugs, learning-based bug detection has emerged only recently.
One example of a learning-based bug detector is DeepBugs, which exploits a kind of information that is typically ignored by program analyses: the implicit information encoded in natural language identifiers.
Due to the inherent fuzziness of this information (see Dimension~1 in Section~\ref{sec:dimensions}) and the fact that determining whether a piece of code is correct is often impossible without a human (Dimension~2), a learning-based approach is a good fit for this problem.
DeepBugs formulates bug detection as a classification problem, i.e., it predicts for a given code snippet whether the code is correct or buggy.

\paragraph{Extracting code examples}
To gather training data, DeepBugs extracts code examples that focus on specific kinds of statements and bug patterns that may arise in these statements.
One of these bug patterns is illustrated at line~\ref{line:bug} of Figure~\ref{fig:example}, where the arguments \code{y} and \code{x} given to a function have been swapped accidentally.
To find such swapped argument bugs, the analysis extracts all function calls with at least two arguments.
Based on the common assumption that most code is correct, the extracted calls serve as examples of correct code.
In contrast to the many correct examples one can extract this way, it is non-obvious how to gather large amounts of incorrect code examples (Dimension~3).
DeepBugs addresses this challenge by artificially introducing bugs into the extracted code examples.
For example, creating swapped argument bugs amounts to simply swapping the arguments of calls found in the code corpus, which is likely to yield incorrect code.
DeepBugs is a generic framework that supports other bug patterns beyond swapped arguments, which are elided here for brevity.

\paragraph{Transformation into vectors}
DeepBugs represents each code example as a concatenation of several pieces of information.
Most importantly, the representation includes the natural language identifiers involved in the code snippet.
For the example bug in Figure~\ref{fig:example}, the analysis extracts the name of the called function, \code{mark\_point}, and the names of the arguments, in particular the two swapped arguments \code{y} and \code{x}.
Beyond identifier names, the analysis also considers contextual information about a code snippet, e.g., the ancestor nodes of the code in the AST or operators involved in an expression.
These pieces of information are represented as vectors based on pre-trained embeddings. 
To represent identifiers, DeepBugs pre-trains a Word2vec model~\cite{Mikolov2013a} on token sequences of source code, which enables the analysis to generalize across similar identifiers. 
For our running example, this generalization allows DeepBugs to understand that when giving arguments named similarly to \code{x} and \code{y} to a function named similarly to \code{mark\_point}, one typically passes \code{x} as the first of the two arguments.

\begin{figure*}
	\includegraphics[width=1\linewidth]{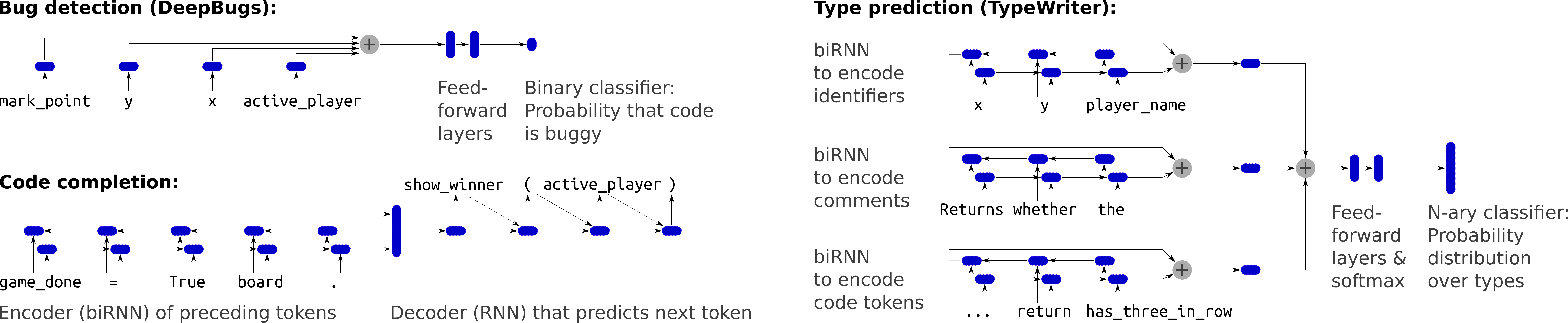}
	\caption{Neural models used in the three example analyses.}
	\label{fig:models examples}
\end{figure*}

\paragraph{Neural model}
The neural model that classifies a given piece of code as buggy or correct is a simple feedforward neural network.
Figure~\ref{fig:models examples} (top-left) illustrates the model with an example.
The model concatenates the embedding vectors of all inputs given to the model and then predicts the probability $p$ that the code is buggy.
For all code examples that are taken from the code corpus without modification, i.e., supposedly correct code, the model is trained to predict $p=0.0$, whereas it is trained to predict $p=1.0$ for the artificially injected bugs.
Once trained, the DeepBugs model can predict for previously unseen code how likely it is that this code is buggy.
To this end, the analysis extracts exactly the same kind of information as during training and queries the classification model with them.
If the model predicts $p$ above some configurable threshold, the analysis reports a warning to the developer.

\paragraph{Validation and ranking}
DeepBugs does not further validate potential bugs before reporting them as warnings to developers.
However, to prioritize warnings, DeepBugs ranks potentially buggy code by the predicted probability $p$.
Developers can then inspect all potential bugs with a $p$ above some threshold and go down the list  starting from the most likely bug.

%

\medskip
\noindent
DeepBugs-inspired code analysis tools are available for various JetBrains IDEs.
Plugins that analyze JavaScript\footnote{\url{https://plugins.jetbrains.com/plugin/12220-deepbugs-for-javascript}} and Python\footnote{\url{https://plugins.jetbrains.com/plugin/12218-deepbugs-for-python}} code have already been downloaded by thousands of developers.

\subsection{Learning to Predict Types}
\label{sec:TypeWriter}

TypeWriter~\cite{fse2020} is a neural software analysis to predict type annotations in dynamically typed languages, such as Python or JavaScript.
While not required in these languages, type annotations are often sought for when projects are growing, as they help ensure correctness, facilitate maintenance, and improve the IDE support available to developers.
To illustrate the problem, consider the \code{mark\_point} function in Figure~\ref{fig:example}.
Since the function does not have any type annotations, the problem is to infer the type of its arguments and its return type.
Type prediction is a prime target for neural software analysis because it fits all three dimensions from Section~\ref{sec:dimensions}.
Source code provides various hints about the type of a variable or function, many of which are fuzzy, such as the name of a variable, the documentation that comes with a function, or how a variable is used (Dimension~1).
Because there sometimes is more than one correct type annotation that a developer could reasonably choose, there is no well-defined criterion to automatically check whether a human will agree with a predicted type (Dimension~2).
Finally, large amounts of code have already been annotated with types, providing sufficient training data for neural models (Dimension~3).

\paragraph{Extracting code examples}
To predict the types of function arguments and return types, TypeWriter extracts two kinds of information from Python code.
On the one hand, the analysis extracts natural language information associated with each possibly type-annotated program element, such as the name of a function argument or a comment associated with the function.
On the other hand, the analysis extracts programming language information, such as how the code element is used.
For example, consider the return type of the \code{mark\_point} function.
TypeWriter extracts the return statement at line~\ref{line:return}, which includes a variable name (\code{has\_...}) and a comment associated with the function (``... Returns whether... ''), which both hint at the return type being Boolean.
Such information is extracted for each function in a given code corpus.
For functions that already have type annotations, the analysis also extracts the existing annotations, which will serve as the ground truth to learn from.

\paragraph{Transformation into vectors}
TypeWriter represents the extracted information as several sequences of vectors based on pre-trained embeddings.
All code tokens and identifier names associated with a type are mapped to vectors using a code token embedding.
The code token embedding is pre-trained on all the Python code that the analysis learns from.
The comments associated with a type are represented as a sequence of words, that each is represented as a vector obtained via a word embedding.
The word embedding is pre-trained on all comments extracted from the given code corpus to ensure that it is well suited for the vocabulary that typically appears in Python comments.

\paragraph{Neural model}
The neural model of TypeWriter (Figure~\ref{fig:models examples}, right) summarizes the given vector sequences using multiple recurrent neural networks:
one for all identifier names associated with the to-be-typed program element,
one for all code tokens related to it, and
one for all related natural language words.
These three recurrent neural networks each result in a vector, and the concatenation of these vectors then serves as input to a feedforward network that acts as a classifier.
The classifier outputs a probability distribution over a fixed set of types, e.g., the 1,000 most common types in the corpus.
TypeWriter interprets the probabilities predicted by the neural model as a ranked list of possible types for the given program element, with the type that has the highest probability at the top of the list.

\paragraph{Validation and ranking}
The top-most predicted type may or may not be correct.
For example, the neural model may predict \code{int} as the type of \code{player\_name}, while it should actually be \code{string}.
To be useful in practice, a type prediction tool must ensure that adding the types it suggests does not introduce any type errors.
TypeWriter uses a gradual type checker to validate the types predicted by the neural model and to find a set of new type annotations that are consistent with each other and with any previously existing annotations.
To this end, the approach turns the problem of assigning one of the predicted types to each of the not yet annotated program elements into a combinatorial search problem.
Guided by the number of type errors that the gradual type checker reports, TypeWriter tries to add as many missing types as possible, without introducing any new type errors.
The type checker-based validation combines the strengths of neural software analysis and traditional program analysis, by using the latter as a validation for the predictions made by the first.

\medskip
\noindent
TypeWriter has been developed at Facebook as a way to add type annotations to Python code.
It has already added several thousands of types to software used by billions of people.

\subsection{Learning to Complete Partial Code}
\label{sec:code completion}

As a third example, we describe a neural code completion technique.
As a developer is typing code in, the technique predicts the next token at a cursor position.
The neural model produces a probability distribution over potential output tokens, and typically, the top five or ten most likely tokens will be shown to the developer.
The next-token prediction problem is highly suited to neural software analysis:
Regarding Dimension~1, the information is fuzzy because there is no easy way to articulate rules that dictate which token should appear next.\footnote{Type information can guide which tokens \emph{cannot} appear next.}
Regarding Dimension~2, there is no well defined correctness criterion, except that the program should continue to compile.
Moreover, the interactivity requirements rule out an expensive validation at each point of prediction.
Finally, regarding Dimension~3, there are copious amounts of training data available, as virtually all code in a given programming language is fair game as training data.

\paragraph{Extracting code examples.}
The code example extraction is trivial:
For any given context---which means the code tokens up to the cursor position---the token that comes immediately after the cursor in a training corpus is the correct prediction.
For the example in Figure~\ref{fig:example}, suppose a developer requests code completion at line~\ref{line:completion} with the cursor at the location marked with \code{???}.
The analysis extracts the sequence of preceding tokens, i.e., $\langle$\code{game\_done}, \code{=}, \code{True}, \code{board}, \code{.}$\rangle$, and the expected prediction here would be \code{show\_winner}.

\paragraph{Transformation into vectors}
Tokens are represented through a fixed-size vocabulary, with out-of-vocabulary tokens being represented by the special ``unknown'' token.
For example, the input above may be re-written to $\langle$42, 233, 8976, 10000, 5$\rangle$, where the numbers are indices into the vocabulary, and \code{board} is an out-of-vocabulary token presented by index 10000.
Also, all input-output examples are typically padded up to be of the same length using an additional padding token.

\paragraph{Neural model}
While predicting the next token seems like a flat classification problem, a typical implementation would often use an encoder-decoder model (Figure~\ref{fig:models examples}, bottom-left).
The first input layer of the encoder is an embedding layer that maps each token into a vector.
The embedding is learned during training in an end-to-end manner, rather than separately pre-trained as in the previous analyses.
After embedding each token, the encoder summarizes the entire sequence into a hidden vector.
In our example analysis, the encoder is rendered using a bi-directional LSTM, i.e., a kind of recurrent neural network that ``reads'' the input sequence both left-to-right and right-to-left.
The decoder is also rendered using a recurrent neural network.
Given the hidden vector, the decoder produces an output sequence, which is expected to equal the input sequence, except shifted by one.
Thus, each step of decoding is expected to produce the next token, considering the context up to that point.
More precisely, the decoder produces at each step a probability distribution over the vocabulary, i.e., the token with maximum probability will be considered as the top-most prediction.
%
During training, the loss is taken point-wise with respect to the ideal decoder output using negative log likelihood loss.
%
As an alternative to the above LSTM-based model, the recently proposed transformer architecture~\cite{Vaswani2017} can also be employed for code prediction~\cite{kim-icse-2021}.

\paragraph{Validation and Ranking}
Code completion in an IDE often shows a ranked list of the most likely next tokens.
Given the probability distribution that the model produces for each token, an IDE can show, e.g., the top five most likely tokens.
Local heuristics, e.g., based on APIs commonly used within the project, may further tweak this ranked list before it is shown to the user.
If developers are interested in predicting more than one token, beam search (Section~\ref{sec:validation ranking}) can predict multiple likely sequences.

\medskip
\noindent
Neural code completion has been an active topic of interest in industry.
For example, it is available in TabNine, studied for internal usage at companies such as Facebook, and for widely-used IDEs, such as IntelliJ from JetBrains and Visual Studio's IntelliCode from Microsoft~\cite{IntelliCompose}.
Recent advances address three problems not considered above.
First, out-of-vocabulary tokens are a crucial problem for code completion because the token to be predicted may not have been seen during training.
The problem can be addressed by splitting complex identifier names into simpler constituent names~\cite{Karampatsis2020a} or by copying tokens from the context using a learned attention mechanism~\cite{li2018code-rnn-attn}.
Second, LSTMs are limited in how much of the code context they remember.
Recent work~\cite{kim-icse-2021, IntelliCompose} addresses this problem through transformer-based architectures, e.g., using a \mbox{GPT-2} transformer model that reasons about a depth-first traversal of the parse tree~\cite{kim-icse-2021}.
Third, the need to predict multiple tokens at a time, e.g., to complete the entire line of code, can be addressed through beam search~\cite{IntelliCompose}.

\section{Outlook and Open Challenges}

Neural software analysis is a fairly recent idea, and researchers and practitioners have just started to explore it.
The following discusses open challenges and gives an outlook into how the field may evolve.

\paragraph{More analysis tasks}
The perhaps most obvious direction for future work is to target more analysis tasks with neural approaches.
In principle, every analysis task can be formulated as a learning problem.
Yet, we see the biggest potential for problems that fit the three dimensions in Section~\ref{sec:dimensions}.
Some tasks that so far have received very little or no attention from the neural software analysis community include the prediction of performance properties of software, automated test input generation, and automated fault injection.

\paragraph{Better ways to gather data}
Most current work focuses on problems for which it is relatively easy to obtain large amounts of high-quality training data.
Once such ``low-hanging fruits'' are harvested, we envision the community to shift attention to more sophisticated ways of obtaining data.
One promising direction is to neurally analyze software based on runtime information.
So far, almost all existing work focuses on static neural software analysis.

\paragraph{Better models}
A core concern of every neural software analysis is how to represent software as vectors that enable a neural model to reason about the software.
Learned representations of code are an active research field with promising  results~\cite{Mou2016,Alon2019,Hellendoorn2020}.
Driven by the observation that code provides a similar degree of ``naturalness'' and regularity as natural language~\cite{Hindle2012}, the community is often driven by techniques that are successful in natural language processing.
Yet, since software and natural language documents are clearly not the same, we envision the focus to move even more toward models specifically designed for software.

\paragraph{Semi-supervised and unsupervised learning}
A promising direction for avoiding the need to obtain labeled training data for supervised learning is semi-supervised and unsupervised learning.
The basic idea is to pre-train a model on some ``pseudo-task'', for which it is easy to obtain large amounts of labeled data, e.g., a language model or an auto-encoder~\cite{Roziere2020}, and to then use the pre-trained model on a related task for which few or even no labeled training data is available.
Such few-to-zero-shot learning shows impressive results on natural language tasks, and is likely to get adopted to software in the future.

\paragraph{Interpretability}
Neural models often suffer from a lack of interpretability.
This general problem affects neural software analysis in particular, because the ``consumers'' of these analyses typically are developers, i.e., human users.
Future research should investigate how to communicate to developers why a model makes a prediction, and how to give developers more confidence in following the predictions of a neural software analysis.
Work on attributing predictions by a model to specific code lines is a promising first step in this direction~\cite{Gupta2019a}.

\paragraph{Integration with traditional program analysis}
Neural and traditional analysis techniques have complementary strengths and weaknesses, and for many problems, combining both may be more effective than each of them individually.
One way of integrating both kinds of analyses could be a continuous feedback loop, where a neural and a traditional analysis repeatedly augment the program in a way that enables the other analysis to cover more cases.
Another idea is to apply traditional program slicing before passing code to a neural analysis, instead of simply passing all code as is typically done today.
TypeWriter (Section~\ref{sec:TypeWriter}) shows an early example of integrating neural and traditional analyses, where the latter validates the predictions made by the former.

\paragraph{Scalability}
Most neural software analyses proposed so far focus on small code snippets, which typically are not larger than a single function.
Future work is likely to tackle the question of how to make predictions about larger pieces of code, e.g., entire modules or applications.
Scaling up neural software analysis will require novel ways of decomposing the reasoning performed by an analysis and of propagating predictions made for one part of a program into another part of the program.

\paragraph{Software-related artifacts beyond the software itself}
This article and most of the neural software analysis work done so far focuses on the software itself.
There are many other artifacts associated with software that could also benefit from neural analysis.
Such artifacts include texts that mix natural language with code elements, e.g., in the communication that happens during code review or in bug reports, and software-generated artifacts, such as crash traces or logs~\cite{issta2020}.

\section{Concluding Remarks}

Neural software analysis is an ambitious idea to address the challenges of a software-dominated world.
The idea has already shown promising results on a variety of development tasks, including work adopted by practitioners.
From a fundamental point of view, neural software analysis provides a powerful and elegant way to reason about the uncertain nature of software.
This uncertainty relates both to the fact that program analysis problems are typically undecidable and to the ``fuzzy'' information, such as natural language and coding conventions, that is embedded into programs.
From a pragmatic point of view, neural software analysis can help in significantly reducing the effort required to produce a software analysis.
While traditional, logic-based analyses are usually built by program analysis experts, of which only a few hundred exist in the world, neural software analyses are learned from data.
This data-driven approach enables millions of developers to, perhaps unconsciously, contribute to the success of neural software analyses by the mere fact that they produce software.
That is, neural software analysis uses the root cause of the demand for developer tools -- the ever-increasing amount, complexity, and diversity of software -- to respond to this demand.

\section*{Acknowledgments}
This work was supported by the European Research Council (ERC, grant agreement 851895), and by the German Research Foundation within the ConcSys and Perf4JS projects.


\bibliographystyle{ACM-Reference-Format}
\bibliography{referencesMP,references}

\end{document}

%% file: lstlistings_python.tex
\makeatletter
\newcommand\HL{%
	\gdef\lst@alloverstyle##1{%
		\fboxrule=0pt
		\fboxsep=0pt
		\colorbox{yellow}{\strut##1}%
	}%
}

\newcommand\HLoff{%
	\xdef\lst@alloverstyle##1{##1}%
}